\documentclass[aip,reprint]{revtex4-1}

\usepackage[export]{adjustbox} 
\usepackage{amsmath}
\usepackage{svg}
\usepackage{graphicx}
\usepackage{dcolumn}
\usepackage{bm}
\usepackage[mathlines]{lineno}
\usepackage[utf8]{inputenc}
\usepackage[T1]{fontenc}
\usepackage{mathptmx}
\usepackage{etoolbox}
\usepackage{float}

\draft 

\makeatletter
\def\@email#1#2{%
 \endgroup
 \patchcmd{\titleblock@produce}
  {\frontmatter@RRAPformat}
  {\frontmatter@RRAPformat{\produce@RRAP{*#1\href{mailto:#2}{#2}}}\frontmatter@RRAPformat}
  {}{}
}%
\makeatother
\begin{document}

\title{Influence of laser chirp and interferometer delay and imbalance on the performance of a time-bin BB84 quantum key distribution system} 

\author{L.Millet}
\affiliation{ID Quantique SA, CH-1227 Genève, Switzerland}
\affiliation{Group of Applied Physics, University of Geneva, CH-1211 Genève, Switzerland}

\author{A.Boaron}
\affiliation{ID Quantique SA, CH-1227 Genève, Switzerland}

\author{R.Thew}
\affiliation{Group of Applied Physics, University of Geneva, CH-1211 Genève, Switzerland}

\author{G.Boso}
\email{gianluca.boso@idquantique.com}
\affiliation{ID Quantique SA, CH-1227 Genève, Switzerland}

\date{\today}

\begin{abstract}
We investigate the effect of interferometer delay and imbalance on the performance of a BB84 time-bin quantum key distribution system. We simulate the impact of interference visibility on system performance and measure the visibility of a pair of interferometers as a function of their relative time delay and intensity imbalance. In addition, our analysis highlights the effect of laser chirp on system performance.

\end{abstract}

\maketitle 

\section{\label{sec:level1}Introduction}
Quantum key distribution (QKD) enables secure, information-theoretic communication through the principles of quantum mechanics, unlike classical cryptographic methods that rely on computational assumptions for their security \cite{gisin2002}. Indeed, in QKD any eavesdropping attempt can be detected through the introduction of errors in the key generation process, whose rate bounds the secret key length \cite{Shor2000,Lim2014}. 

In BB84\cite{bennett1984}, the first and most widely used QKD protocol, the secret key is typically encoded in either the computational (Z) or Hadamard (X) basis. The phase error rate ($PER$) of the qubits encoding the key is particularly important because it bounds the fraction of secure bits, but is not directly measurable. Instead, the $PER$ in the Z (X) basis is estimated from the quantum bit error rate in the X (Z) basis ($QBER$), as a bit-flip error in one basis corresponds to a phase-flip in the other\cite{fung2010}.

However, errors do not only originate from a potential eavesdropper but also from imperfections in the physical QKD setup\cite{fadri2020apl,xu2020,Currás2024}. In a typical time-bin BB84 protocol, Alice’s side uses an imbalanced interferometer to create pairs of phase-correlated weak coherent pulses, encoding qubits in either the Z or X basis using intensity and phase modulators. At Bob’s end, after either a passive or an active basis-choice mechanism, Z-basis qubits are directly detected by a single-photon detector and are used to encode the key. Meanwhile, X-basis qubits interfere within a matching interferometer for measurement, and its quantum bit error rate ($QBER_X$) is used to estimate the $PER$ in the Z basis. The alignment of these two interferometers, notably their temporal and intensity matching, is critical for reducing the $QBER_X$. In fact, interference visibility $V$ correlates with the quantum bit errors in the X basis, according to $V=1-2QBER_X$\cite{gisin2002}. 

In this work, we explore how the performance of a BB84 QKD time-bin system is influenced by the time and intensity mismatch between the two interferometers. To do so, we first simulate the impact of visibility on the performance of a time-bin BB84 QKD system. We then measure the visibility of a pair of Michelson interferometers as a function of their intensity imbalance and relative time delay. We also investigate the impact of laser chirp, variations in the laser pulse wavelength over time, on the performance of the QKD system. We use an analytical model to show the effects of chirp on visibility in the presence of imbalance and delay and compare it with the unchirped case. Finally, we measure the chirp of our optical source and highlight its impact on visibility.

\section{\label{sec:level1}Impact of visibility on the performance of a time-bin BB84 protocol}

To quantify the impact of the visibilities of imperfect interferometers on QKD performance, we simulate a QKD system based on a 2-decoy state time-bin and phase BB84 working at a qubit repetition rate of 500\,MHz\cite{Lim2014}.
The single-photon detectors have a dark count rate (DCR) of 1000\,Hz.
We vary the quality of the state preparation and measurement in the X basis and calculate the secret key rate (SKR) as a function of the attenuation between Alice and Bob.
We analyze the performance of the system as a function of the system visibility at 0\,dB, which is defined as $V =1-2 m_X / n_X$\cite{Lim2014}, where $n_X$ and $m_X$ are the numbers of events and errors, respectively, which occur in the X basis.
This quantity is primarily affected by the visibility of the interferometer pair because the contribution of the detectors DCR at 0\,dB is almost negligible.

\begin{figure}
    \centering
    \includegraphics[width=\columnwidth, trim=0mm 0mm 0mm 0mm, clip]{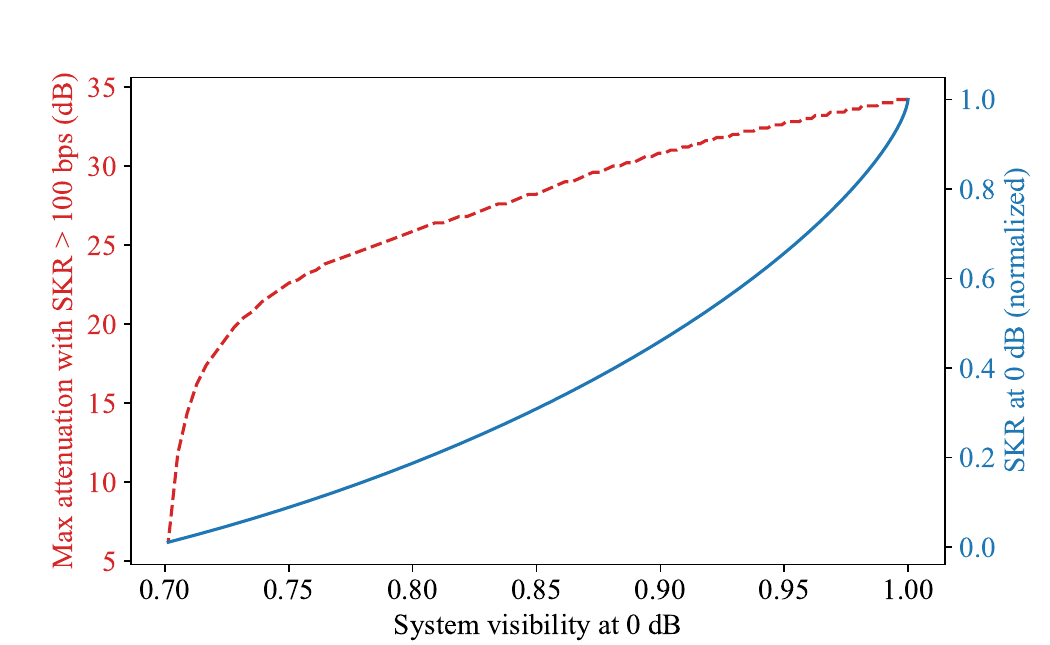}
    \caption{Maximum attenuation achievable with a secret key rate (SKR) higher than 100 bps (red dashed line) and  SKR at 0\,dB normalized by the SKR achieved with a visibility of 1 (blue line) as a function of the system visibility at 0\,dB.}
    \label{fig:system_perf}
\end{figure}

We use two figures of merit: firstly, the SKR at 0\,dB normalized by the SKR achieved with a visibility of 1; secondly, the maximum attenuation achievable with a SKR higher than 100\,bps.
They are represented in Fig.\,\ref{fig:system_perf} as a function of the system visibility at 0\,dB.
The performance of the system degrades steadily when visibility decreases, and the SKR drops to nearly 0 at 0.70 visibility.
Therefore, there is no trade-off to be done on the system visibility: one should always seek the highest visibility.
Compared to the maximum visibility, a visibility of 0.98 leads to a SKR at 0\,dB reduced by 19\,\% and a maximum attenuation reduced by 0.6\,dB.

\section{\label{sec:level1}Measuring the visibility}
\begin{figure}
    \centering
    \includegraphics[width=\columnwidth, trim=0mm 0mm 12mm 0mm, clip]{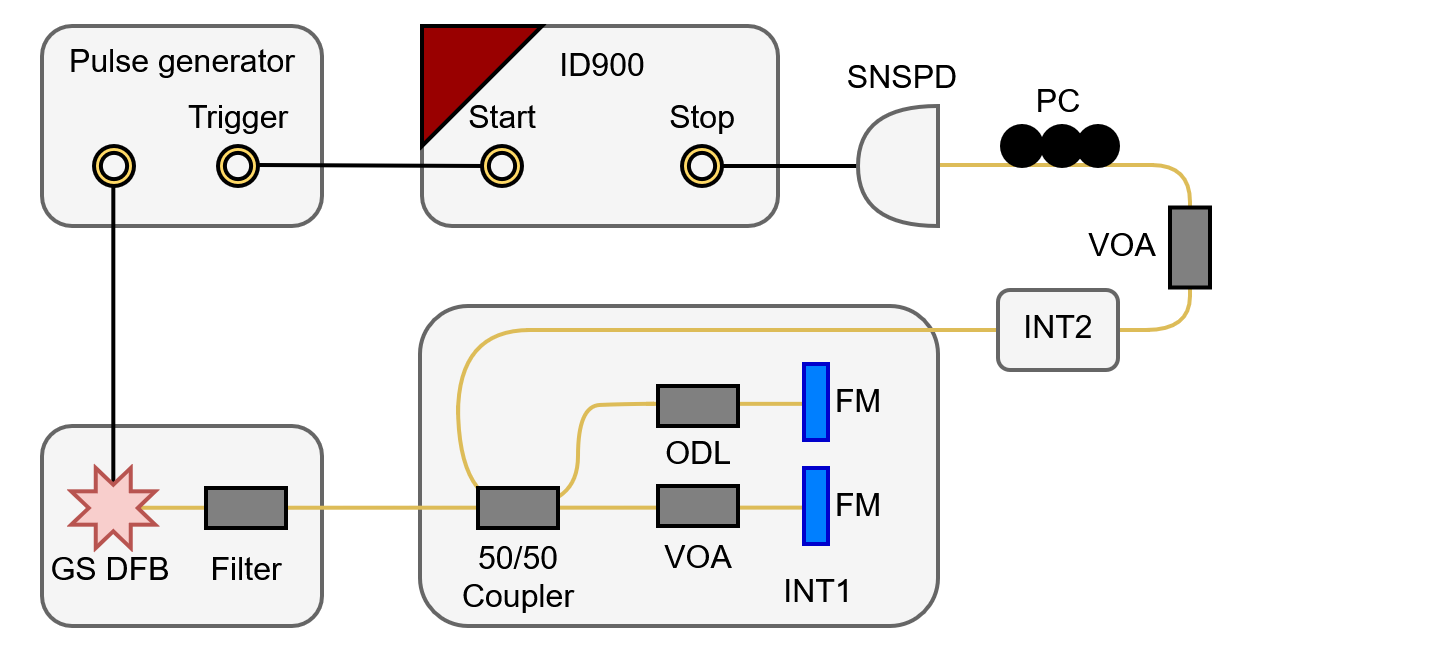} % Scale to column width
    \caption{\label{fig:epsart} Schematic of the visibility measurement setup. GS DFB: gain-switched distributed feedback laser. INT: interferometer. ODL: optical delay line. VOA: variable optical attenuator. FM: Faraday mirror. PC: polarization controller. SNSPD: superconducting nanowire single photon detector.}
    \hspace{8mm}
    \label{fig:visibility_setup}
\end{figure}

The experimental setup to measure the visibility of a pair of interferometers as a function of delay and intensity imbalance is illustrated in Fig.\,\ref{fig:visibility_setup}. We use a commercially available 1550\,nm Distributed Feedback (DFB) laser diode pulsed at a 500\,MHz repetition rate, followed by a fiber-based band-pass filter of 114\,pm FWHM at 3\,dB, resulting in a 77\,picosecond FWHM pulse. Laser pulses are sent through a fiber Michelson interferometer (INT1) featuring tunable delay (± 11\,ps) via a mechanical optical delay line (ODL) and adjustable imbalance through an electrical variable optical attenuator (VOA). Faraday mirrors (FM) ensure polarization insensitivity. This interferometer generates phase-correlated pulse pairs and allows us to sweep the delay and imbalance. 

The pulses then reach a second Michelson interferometer (INT2), which has a fixed 1\,ns delay and negligible intensity imbalance between its arms. Interference occurs between the pulses that traverse the short arm of INT1 and the long arm of INT2 and those taking the long arm of INT1 and the short arm of INT2. 

Detection is performed using a superconducting nanowire single photon
detector (ID Quantique, ID 281). A polarization controller (PC) preceeds the SNSPD to optimize its detection efficiency. Intensity measurements are extracted from a Time-Correlated Single-Photon Counting (TCSPC) histogram, acquired by a start-stop time to digital converter (ID Quantique, ID900).

We define the intensity imbalance of a given interferometer as $U=(I_{short}-I_{long})/\operatorname{max}[(I_{short},I_{long})]$, with $I_{short}=\eta I_{long}$, and $I$ the pulse intensity. The imbalance of the pair of interferometers is approximated by $U_{INT1}$, as $U_{INT2}$ is negligible. For a given delay, we tune the VOA voltage to set $U_{INT1}$ at specific values between ± 44\,\%. 
To maximize the intensity of constructive interferences ($I_{max}$) and minimize that of destructive interferences ($I_{min}$), the relative phase within pulse pairs is optimized by fine-tuning the temperature of INT2. The visibility of the interference is then calculated using the formula: $V=(I_{max}-I_{min})⁄(I_{max}+I_{min})$.

\section{\label{sec:level1}Experimental results }
\begin{figure}
    \centering
    \includegraphics[width=\columnwidth]{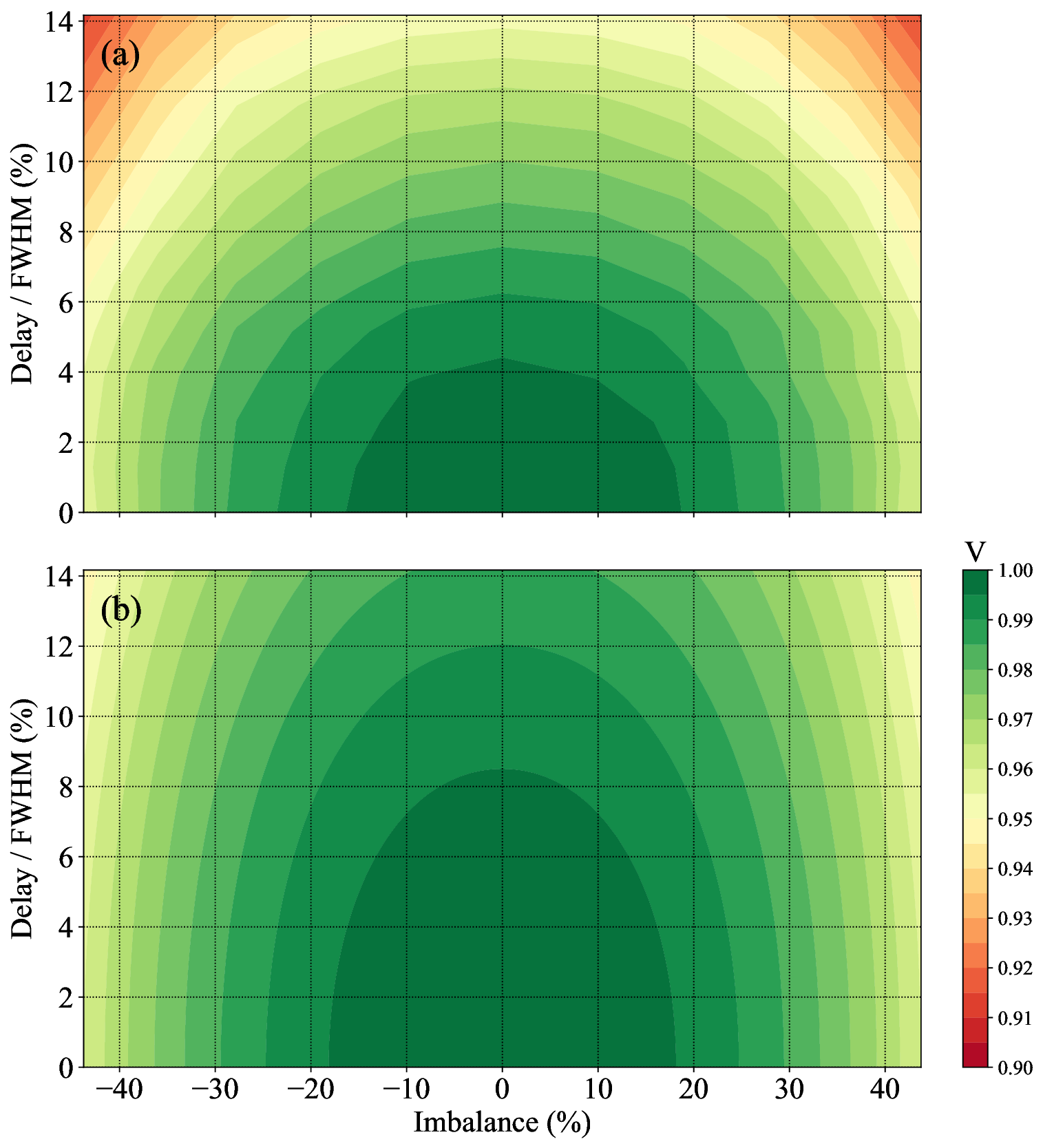} % Scale to column width
    \caption{\label{fig:epsart} (a) Experimental visibility contour plot as a function of imbalance and delay. (b) Analytically calculated visibility contour plot of an unchirped Gaussian laser pulse.}
    \label{fig:exp_visibility_plot}
\end{figure}

We measure the visibility for delays ranging from 0 to 11\,picosecond, with a 0\,ps delay indicating an equal delay between the two arms of INT1 and INT2, and a total imbalance ranging from ± 44\,\% (see Fig.\,\ref{fig:exp_visibility_plot}). To allow comparison across systems with different pulse durations, the vertical axis is normalized as the ratio between the delay and the pulse FWHM; for instance, an 11\,ps delay corresponds to approximately 14\,\% of the FWHM. The maximum measured visibility at 0 delay and imbalance was 99.97\,\%, which is very close to the theoretical maximum. As expected, visibility decreases with increasing delay and imbalance values, reaching a minimum of approximately 91\,\% at 44\,\% imbalance and 14\,\% of normalized delay.

\section{\label{sec:level1}Laser chirp impact on performance}
To assess how our experiment deviates from the ideal case, we analytically calculate the visibility of two interfering Gaussian pulses. The electric field of the first pulse is represented as $E_1(t)=\sqrt{I_1(t)}\cos(\omega_{0}t+\phi(t))$ ,  while the second pulse, delayed by a time $\tau$, is represented as $E_2=\sqrt{\eta I_1(t)}\cos(\omega_{0}(t-\tau)+\phi(t-\tau))$ . 

$I(t)$ is the Gaussian intensity of a 77\,ps pulse, and the parameter $\eta$ quantifies the attenuation of the second pulse, defined according to the intensity imbalance described earlier. $\omega_0$ is the pulse carrier angular frequency, and $\phi(t)$ is a time-dependent phase. The chirp is defined as $\Delta\omega \equiv \omega_{inst}-\omega_0=d\phi(t)/dt$, with $\omega_{inst}$ the instantaneous angular frequency of the pulse. We compute visibility as a function of delay and imbalance using the following analytical expression \cite{steck2007}:
\begin{equation}
    V={\frac{2\sqrt{\eta}}{1+\eta}}| g^{(1)}(\tau) |\,,
\end{equation}
where $| g^{(1)}(\tau) |$ is the magnitude of the correlation function. For unchirped Gaussian pulses ($\phi(t)=0$), the correlation function becomes $| g^{(1)}(\tau) | = \exp({-1/2\left({\tau}/{\text{t}_g}\right)^2)}$, with $t_g = {\text{FWHM}}/\sqrt{2 \ln 2}$. The calculated visibility ($V_{sim}$) of unchirped Gaussian pulses is significantly higher than experimental visibility ($V_{exp}$) (see Fig.\,\ref{fig:exp_visibility_plot}). To quantify the discrepancy, we define for a given delay and imbalance an error $\epsilon$ as:
\begin{equation}
    \epsilon = \frac{V_{sim}-V_{exp}}{1-V_{exp}}\times100\,
\end{equation}
On average, $\epsilon$ was 34.3\,\%. This discrepancy points to the substantial degradation in visibility caused by chirp, emphasizing the need to mitigate chirp in DFB lasers to improve the interferometer performance of QKD systems. 
For example, at 0\,\% imbalance and 10\,\% normalized delay, visibility decreases from approximately 0.993 (no chirp) to 0.975 (with chirp)  corresponding to a 22\,\% reduction in the SKR at 0 \,dB instead of a 9\,\% reduction.

To further validate the origin of the discrepancy, we measure the chirp of our optical source and simulate the visibility based on the retrieved phase and measured intensity profile.

\section{\label{sec:level1}Chirp measurement}
We used the phase reconstruction by ultrafast optical differentiation method  \cite{Li2009} (PROUD) to retrieve the chirp. PROUD determines the chirp by comparing two temporal intensity profiles: one from the original pulse and the second after propagation through a positive or negative linear frequency filter. Because frequency filtering performs time derivation of the propagating pulse, one can retrieve the chirp unambiguously from a simple formula: 
\begin{equation}
    \frac{d\phi(t)}{dt}=\pm\sqrt{\frac{\frac{I_{\pm} (t)}{A^2} - (\frac{d|s(t)|}{dt})^2 }{|s(t)|^2 }}\mp\Delta\omega\,,
\end{equation}
where $|s(t)|^2$ is the pulse intensity profile, $I_{\pm} (t)$ is the intensity profile of the positively or negatively filtered pulse, $A$ is the slope of the filter’s transfer function magnitude, and $\Delta\omega$ is the absolute difference between the laser carrier angular frequency and the filter resonance angular frequency. 

\begin{figure}
    \hspace{-8mm}
    \centering
    \includegraphics[width=\columnwidth]{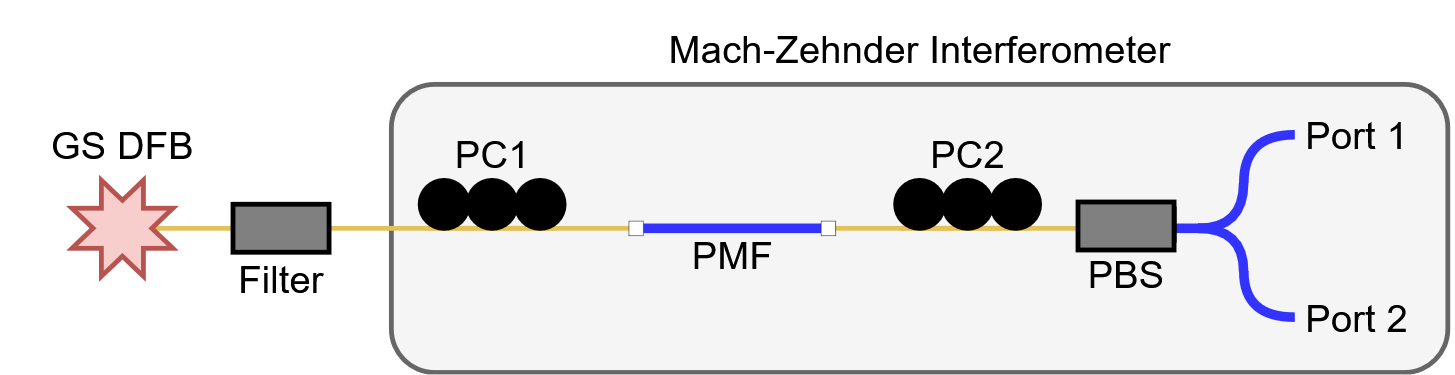} % Scale to column width
    \caption{\label{fig:epsart} Schematic of the PROUD measurement setup using a polarization Mach-Zehnder interferometer. GS DFB: gain-switched distributed feedback laser. PC: polarization controller. PMF: polarization maintaining fiber. PBS: polarizing beam splitter.}
    \label{fig:PROUD_setup}
\end{figure}

We implement linear frequency filtering using a polarization interferometer based on a polarization-maintaining fiber \cite{Consoli2011} (PMF). A key advantage of this PROUD setup is its inherent synchronization between the unfiltered and filtered pulses, eliminating the need for delay compensation. However, our setup differs in the use of a polarizing beam splitter (PBS) placed at the output of the second PC (see Fig.\,\ref{fig:PROUD_setup}). In this configuration, the two output ports of the PBS act as the arms of a Mach-Zehnder interferometer enabling chirp measurements with the filtered intensities from both arms, using the following equation\cite{Li_MZI_2009}:
\begin{equation}
    \frac{d\phi(t)}{dt}=\frac{I_+ (t)-I_- (t)}{4A^2 |s(t)|^2 \Delta\omega}\,,
\end{equation}

The recovered chirp along with the measured pulse intensity profile are shown in Fig.\,\ref{fig:chirp_plot}, and lead to a 14 \% error in reconstructing the spectrum \cite{Consoli2011} (See Supplementary Materials for details about the measurement).

\begin{figure}
    \vspace{5mm}
    \includegraphics[width=\columnwidth, height = 8cm, keepaspectratio]{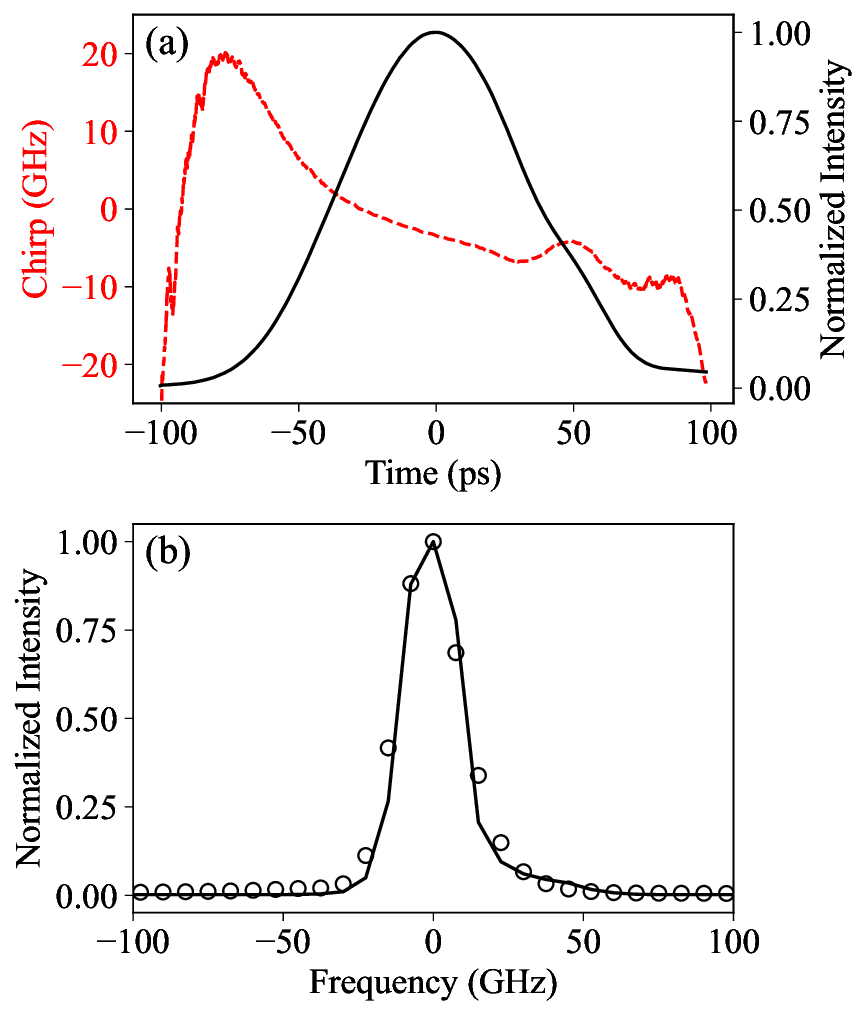} 
    \caption{\label{fig:epsart} (a) Measured chirp (red dashed line) and intensity (black continuous line) of the optical source. (b) Comparison between the reconstructed spectrum (black dots) and measured spectrum (black continuous line) of the optical source.}
    \label{fig:chirp_plot}
\end{figure}

\begin{figure*}
    \includegraphics[height = 5.5cm, keepaspectratio]{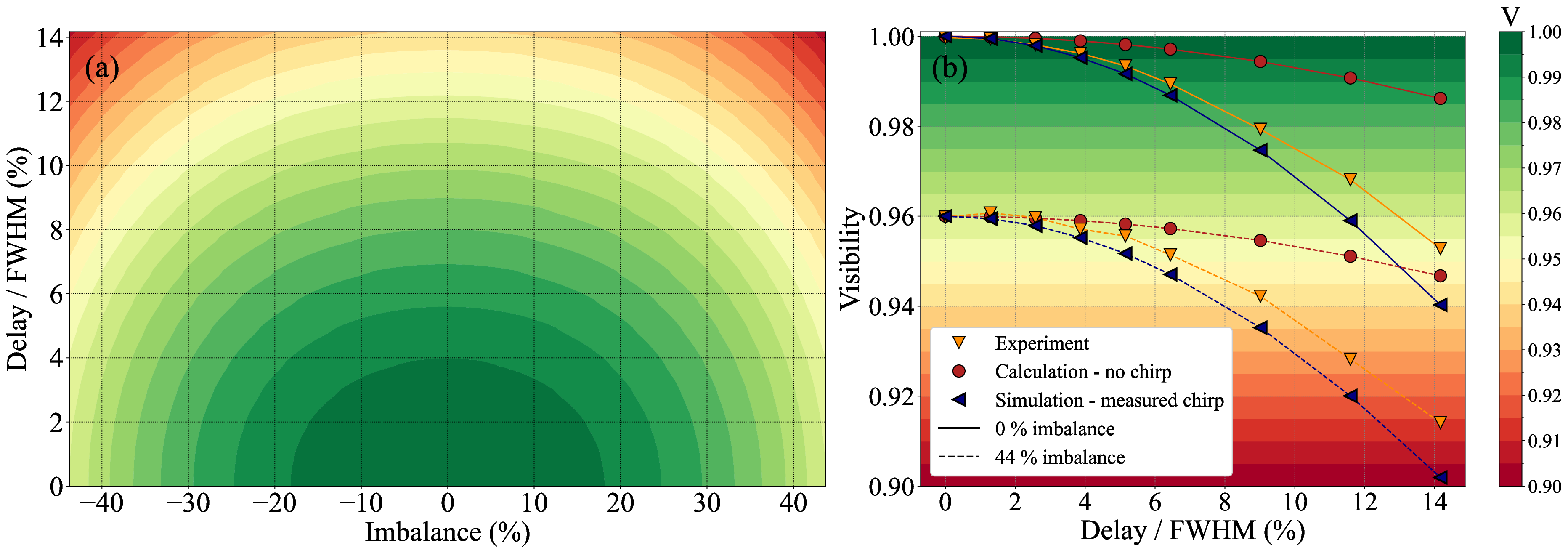} 
    \caption{\label{fig:wide} (a) Simulated visibility contour plot using the measured chirp and intensity profile. (b) Slice at 44 \% and 0 \% of intensity imbalance.}
    \label{fig:visibility_end}
\end{figure*}

Visibility simulations using the retrieved chirp and measured intensity profile yield results similar to those obtained in the experiment. As shown in Fig.\,\ref{fig:visibility_end}, chirp does not affect the visibility of interfering pulses at zero delay, but its impact becomes significant under nonzero delay conditions. On average, $\epsilon$ is now 13.9\,\%, demonstrating that the measured chirp can explain a significant proportion of discrepancies.

Finally, we extend the analytical calculation by modeling the time-dependent phase $\phi(t)$ using a Taylor expansion: $\phi(t)=\phi_0+\alpha t+ \beta t^2+\ldots$, ensuring that the resulting visibility fits the experimental data. We considered only the second-order term $\beta$ of the Taylor expansion as nonzero, acknowledging that chirp in DFB lasers is predominantly linear around the peak intensity \cite{Shakhovoy2021}, and lower-order terms contribute to a constant phase offset that does not affect the visibility. For linearly chirped Gaussian pulses, the magnitude of the correlation function becomes:
\begin{equation}
    | g^{(1)}(\tau) | = \exp(-{1/2\left(\frac{\tau}{\text{t}_g} \right)^2(1+\beta^2{\text{t}_g}^4))},
\end{equation}
This expression indicates that the presence of chirp reduces visibility by a factor proportional to $\beta^2{\text{t}_g}^4$. A fitted value of  $\beta \approx -3.7\times10^{20}$ $\operatorname{rad.s^{-2}}$  (Fig. S2) is used to match the experimental visibility and leads to an average $\epsilon$ of 5.9\,\% (Fig. S3). This lowest error value suggests that the fitted temporal phase provides a more accurate representation of the phase of our optical source, as the higher error using the measured chirp can be attributed to the accuracy limitations of our PROUD setup. 

\section{\label{sec:level1}Discussion}
Interferometers are a critical component of QKD systems, as their visibility directly affects both the SKR and the maximum achievable distance. For instance, a 2\,\% reduction in visibility from 1 to 0.98 already leads to a 19\,\% decrease in SKR for the protocol considered here. This dependence shows the importance of carefully managing all the parameters that influence visibility. Indeed, we have shown how the relative delay and intensity imbalance between interferometers degrade visibility, and that the degradation is aggravated in the presence of chirp at nonzero delays. For linearly chirped Gaussian pulses, the visibility is reduced by a factor of $\exp(-\frac{1}{2} \left( \frac{\tau}{\text{t}_g} \right)^2 \beta^2 \text{t}_g^4)$, which typically reduces the expected SKR by an additional 10\,\% given the chirp of our optical source. Our findings underscore the importance of minimizing both the interferometric delay and imbalance and of reducing laser chirp through appropriate source selection or chirp compensation schemes. These optimizations are essential for maximizing visibility and thereby improving QKD performance.

\vspace{5\baselineskip}
\section*{Supplementary Material}
See Supplementary Materials for further explanations about chirp measurement and visibility simulations. 

\begin{acknowledgments}
The authors acknowledge financial support from the Marie Sklodowska-Curie Grant No. 101072637 (Project Quantum-Safe-Internet).
\end{acknowledgments}

%\bibliography{IF_BB84}

\begin{thebibliography}{13}%
\makeatletter
\providecommand \@ifxundefined [1]{%
 \@ifx{#1\undefined}
}%
\providecommand \@ifnum [1]{%
 \ifnum #1\expandafter \@firstoftwo
 \else \expandafter \@secondoftwo
 \fi
}%
\providecommand \@ifx [1]{%
 \ifx #1\expandafter \@firstoftwo
 \else \expandafter \@secondoftwo
 \fi
}%
\providecommand \natexlab [1]{#1}%
\providecommand \enquote  [1]{``#1''}%
\providecommand \bibnamefont  [1]{#1}%
\providecommand \bibfnamefont [1]{#1}%
\providecommand \citenamefont [1]{#1}%
\providecommand \href@noop [0]{\@secondoftwo}%
\providecommand \href [0]{\begingroup \@sanitize@url \@href}%
\providecommand \@href[1]{\@@startlink{#1}\@@href}%
\providecommand \@@href[1]{\endgroup#1\@@endlink}%
\providecommand \@sanitize@url [0]{\catcode `\\12\catcode `\$12\catcode `\&12\catcode `\#12\catcode `\^12\catcode `\_12\catcode `\%12\relax}%
\providecommand \@@startlink[1]{}%
\providecommand \@@endlink[0]{}%
\providecommand \url  [0]{\begingroup\@sanitize@url \@url }%
\providecommand \@url [1]{\endgroup\@href {#1}{\urlprefix }}%
\providecommand \urlprefix  [0]{URL }%
\providecommand \Eprint [0]{\href }%
\providecommand \doibase [0]{http://dx.doi.org/}%
\providecommand \selectlanguage [0]{\@gobble}%
\providecommand \bibinfo  [0]{\@secondoftwo}%
\providecommand \bibfield  [0]{\@secondoftwo}%
\providecommand \translation [1]{[#1]}%
\providecommand \BibitemOpen [0]{}%
\providecommand \bibitemStop [0]{}%
\providecommand \bibitemNoStop [0]{.\EOS\space}%
\providecommand \EOS [0]{\spacefactor3000\relax}%
\providecommand \BibitemShut  [1]{\csname bibitem#1\endcsname}%
\let\auto@bib@innerbib\@empty
%</preamble>
\bibitem [{\citenamefont {Gisin}\ \emph {et~al.}(2002)\citenamefont {Gisin}, \citenamefont {Ribordy}, \citenamefont {Tittel},\ and\ \citenamefont {Zbinden}}]{gisin2002}%
  \BibitemOpen
  \bibfield  {author} {\bibinfo {author} {\bibfnamefont {N.}~\bibnamefont {Gisin}}, \bibinfo {author} {\bibfnamefont {G.}~\bibnamefont {Ribordy}}, \bibinfo {author} {\bibfnamefont {W.}~\bibnamefont {Tittel}}, \ and\ \bibinfo {author} {\bibfnamefont {H.}~\bibnamefont {Zbinden}},\ }\bibfield  {title} {\enquote {\bibinfo {title} {Quantum cryptography},}\ }\href@noop {} {\bibfield  {journal} {\bibinfo  {journal} {Rev. Mod. Phys.}\ }\textbf {\bibinfo {volume} {74}},\ \bibinfo {pages} {145--195} (\bibinfo {year} {2002})}\BibitemShut {NoStop}%
\bibitem [{\citenamefont {Shor}\ and\ \citenamefont {Preskill}(2000)}]{Shor2000}%
  \BibitemOpen
  \bibfield  {author} {\bibinfo {author} {\bibfnamefont {P.~W.}\ \bibnamefont {Shor}}\ and\ \bibinfo {author} {\bibfnamefont {J.}~\bibnamefont {Preskill}},\ }\bibfield  {title} {\enquote {\bibinfo {title} {Simple proof of security of the bb84 quantum key distribution protocol},}\ }\href@noop {} {\bibfield  {journal} {\bibinfo  {journal} {Phys. Rev. Lett.}\ }\textbf {\bibinfo {volume} {85}},\ \bibinfo {pages} {441--444} (\bibinfo {year} {2000})}\BibitemShut {NoStop}%
\bibitem [{\citenamefont {Lim}\ \emph {et~al.}(2014)\citenamefont {Lim}, \citenamefont {Curty}, \citenamefont {Walenta}, \citenamefont {Xu},\ and\ \citenamefont {Zbinden}}]{Lim2014}%
  \BibitemOpen
  \bibfield  {author} {\bibinfo {author} {\bibfnamefont {C.~C.~W.}\ \bibnamefont {Lim}}, \bibinfo {author} {\bibfnamefont {M.}~\bibnamefont {Curty}}, \bibinfo {author} {\bibfnamefont {N.}~\bibnamefont {Walenta}}, \bibinfo {author} {\bibfnamefont {F.}~\bibnamefont {Xu}}, \ and\ \bibinfo {author} {\bibfnamefont {H.}~\bibnamefont {Zbinden}},\ }\bibfield  {title} {\enquote {\bibinfo {title} {Concise security bounds for practical decoy-state quantum key distribution},}\ }\href {\doibase 10.1103/PhysRevA.89.022307} {\bibfield  {journal} {\bibinfo  {journal} {Phys. Rev. A}\ }\textbf {\bibinfo {volume} {89}},\ \bibinfo {pages} {022307} (\bibinfo {year} {2014})}\BibitemShut {NoStop}%
\bibitem [{\citenamefont {Bennett}\ and\ \citenamefont {Brassard}(1984)}]{bennett1984}%
  \BibitemOpen
  \bibfield  {author} {\bibinfo {author} {\bibfnamefont {C.~H.}\ \bibnamefont {Bennett}}\ and\ \bibinfo {author} {\bibfnamefont {G.}~\bibnamefont {Brassard}},\ }\bibfield  {title} {\enquote {\bibinfo {title} {Quantum cryptography: Public key distribution and coin tossing},}\ }\href@noop {} {\bibfield  {journal} {\bibinfo  {journal} {Proceedings of IEEE International Conference on Computers, Systems and Signal Processing}\ ,\ \bibinfo {pages} {175--179}} (\bibinfo {year} {1984})}\BibitemShut {NoStop}%
\bibitem [{\citenamefont {Fung}, \citenamefont {Ma},\ and\ \citenamefont {Chau}(2010)}]{fung2010}%
  \BibitemOpen
  \bibfield  {author} {\bibinfo {author} {\bibfnamefont {C.-H.~F.}\ \bibnamefont {Fung}}, \bibinfo {author} {\bibfnamefont {X.}~\bibnamefont {Ma}}, \ and\ \bibinfo {author} {\bibfnamefont {H.~F.}\ \bibnamefont {Chau}},\ }\bibfield  {title} {\enquote {\bibinfo {title} {Practical issues in quantum-key-distribution postprocessing},}\ }\href {\doibase 10.1103/PhysRevA.81.012318} {\bibfield  {journal} {\bibinfo  {journal} {Physical Review A}\ }\textbf {\bibinfo {volume} {81}},\ \bibinfo {pages} {012318} (\bibinfo {year} {2010})}\BibitemShut {NoStop}%
\bibitem [{\citenamefont {Grünenfelder}\ \emph {et~al.}(2020)\citenamefont {Grünenfelder}, \citenamefont {Boaron}, \citenamefont {Rusca}, \citenamefont {Martin},\ and\ \citenamefont {Zbinden}}]{fadri2020apl}%
  \BibitemOpen
  \bibfield  {author} {\bibinfo {author} {\bibfnamefont {F.}~\bibnamefont {Grünenfelder}}, \bibinfo {author} {\bibfnamefont {A.}~\bibnamefont {Boaron}}, \bibinfo {author} {\bibfnamefont {D.}~\bibnamefont {Rusca}}, \bibinfo {author} {\bibfnamefont {A.}~\bibnamefont {Martin}}, \ and\ \bibinfo {author} {\bibfnamefont {H.}~\bibnamefont {Zbinden}},\ }\bibfield  {title} {\enquote {\bibinfo {title} {Performance and security of 5 ghz repetition rate polarization-based quantum key distribution},}\ }\href@noop {} {\bibfield  {journal} {\bibinfo  {journal} {Applied Physics Letters}\ }\textbf {\bibinfo {volume} {117}},\ \bibinfo {pages} {144003} (\bibinfo {year} {2020})}\BibitemShut {NoStop}%
\bibitem [{\citenamefont {Xu}\ \emph {et~al.}(2020)\citenamefont {Xu}, \citenamefont {Ma}, \citenamefont {Zhang}, \citenamefont {Lo},\ and\ \citenamefont {Pan}}]{xu2020}%
  \BibitemOpen
  \bibfield  {author} {\bibinfo {author} {\bibfnamefont {F.}~\bibnamefont {Xu}}, \bibinfo {author} {\bibfnamefont {X.}~\bibnamefont {Ma}}, \bibinfo {author} {\bibfnamefont {Q.}~\bibnamefont {Zhang}}, \bibinfo {author} {\bibfnamefont {H.-K.}\ \bibnamefont {Lo}}, \ and\ \bibinfo {author} {\bibfnamefont {J.-W.}\ \bibnamefont {Pan}},\ }\bibfield  {title} {\enquote {\bibinfo {title} {Secure quantum key distribution with realistic devices},}\ }\href@noop {} {\bibfield  {journal} {\bibinfo  {journal} {Reviews of modern physics}\ }\textbf {\bibinfo {volume} {92}},\ \bibinfo {pages} {025002} (\bibinfo {year} {2020})}\BibitemShut {NoStop}%
\bibitem [{\citenamefont {Currás-Lorenzo}\ \emph {et~al.}(2023)\citenamefont {Currás-Lorenzo}, \citenamefont {Nahar}, \citenamefont {Lütkenhaus}, \citenamefont {Tamaki},\ and\ \citenamefont {Curty}}]{Currás2024}%
  \BibitemOpen
  \bibfield  {author} {\bibinfo {author} {\bibfnamefont {G.}~\bibnamefont {Currás-Lorenzo}}, \bibinfo {author} {\bibfnamefont {S.}~\bibnamefont {Nahar}}, \bibinfo {author} {\bibfnamefont {N.}~\bibnamefont {Lütkenhaus}}, \bibinfo {author} {\bibfnamefont {K.}~\bibnamefont {Tamaki}}, \ and\ \bibinfo {author} {\bibfnamefont {M.}~\bibnamefont {Curty}},\ }\bibfield  {title} {\enquote {\bibinfo {title} {Security of quantum key distribution with imperfect phase randomisation},}\ }\href@noop {} {\bibfield  {journal} {\bibinfo  {journal} {Quantum Science and Technology}\ }\textbf {\bibinfo {volume} {9}},\ \bibinfo {pages} {015025} (\bibinfo {year} {2023})}\BibitemShut {NoStop}%
\bibitem [{\citenamefont {Steck}(2007)}]{steck2007}%
  \BibitemOpen
  \bibfield  {author} {\bibinfo {author} {\bibfnamefont {D.~A.}\ \bibnamefont {Steck}},\ }\bibfield  {title} {\enquote {\bibinfo {title} {Quantum and atom optics},}\ }\href@noop {} {\  (\bibinfo {year} {2007})}\BibitemShut {NoStop}%
\bibitem [{\citenamefont {Li}, \citenamefont {Park},\ and\ \citenamefont {{n}a}(2009{\natexlab{a}})}]{Li2009}%
  \BibitemOpen
  \bibfield  {author} {\bibinfo {author} {\bibfnamefont {F.}~\bibnamefont {Li}}, \bibinfo {author} {\bibfnamefont {Y.}~\bibnamefont {Park}}, \ and\ \bibinfo {author} {\bibfnamefont {J.~A.}\ \bibnamefont {{n}a}},\ }\bibfield  {title} {\enquote {\bibinfo {title} {Linear characterization of optical pulses with durations ranging from the picosecond to the nanosecond regime using ultrafast photonic differentiation},}\ }\href@noop {} {\bibfield  {journal} {\bibinfo  {journal} {J. Lightwave Technol.}\ }\textbf {\bibinfo {volume} {27}},\ \bibinfo {pages} {4623--4633} (\bibinfo {year} {2009}{\natexlab{a}})}\BibitemShut {NoStop}%
\bibitem [{\citenamefont {Consoli}, \citenamefont {Tijero},\ and\ \citenamefont {Esquivias}(2011)}]{Consoli2011}%
  \BibitemOpen
  \bibfield  {author} {\bibinfo {author} {\bibfnamefont {A.}~\bibnamefont {Consoli}}, \bibinfo {author} {\bibfnamefont {J.~M.~G.}\ \bibnamefont {Tijero}}, \ and\ \bibinfo {author} {\bibfnamefont {I.}~\bibnamefont {Esquivias}},\ }\bibfield  {title} {\enquote {\bibinfo {title} {Time resolved chirp measurements of gain switched semiconductor laser using a polarization based optical differentiator},}\ }\href@noop {} {\bibfield  {journal} {\bibinfo  {journal} {Opt. Express}\ }\textbf {\bibinfo {volume} {19}},\ \bibinfo {pages} {10805--10812} (\bibinfo {year} {2011})}\BibitemShut {NoStop}%
\bibitem [{\citenamefont {Li}, \citenamefont {Park},\ and\ \citenamefont {{n}a}(2009{\natexlab{b}})}]{Li_MZI_2009}%
  \BibitemOpen
  \bibfield  {author} {\bibinfo {author} {\bibfnamefont {F.}~\bibnamefont {Li}}, \bibinfo {author} {\bibfnamefont {Y.}~\bibnamefont {Park}}, \ and\ \bibinfo {author} {\bibfnamefont {J.~A.}\ \bibnamefont {{n}a}},\ }\bibfield  {title} {\enquote {\bibinfo {title} {Single-shot real-time frequency chirp characterization of telecommunication optical signals based on balanced temporal optical differentiation},}\ }\href@noop {} {\bibfield  {journal} {\bibinfo  {journal} {Opt. Lett.}\ }\textbf {\bibinfo {volume} {34}},\ \bibinfo {pages} {2742--2744} (\bibinfo {year} {2009}{\natexlab{b}})}\BibitemShut {NoStop}%
\bibitem [{\citenamefont {Shakhovoy}\ \emph {et~al.}(2021)\citenamefont {Shakhovoy}, \citenamefont {Sharoglazova}, \citenamefont {Udaltsov}, \citenamefont {Duplinskiy}, \citenamefont {Kurochkin},\ and\ \citenamefont {Kurochkin}}]{Shakhovoy2021}%
  \BibitemOpen
  \bibfield  {author} {\bibinfo {author} {\bibfnamefont {R.}~\bibnamefont {Shakhovoy}}, \bibinfo {author} {\bibfnamefont {V.}~\bibnamefont {Sharoglazova}}, \bibinfo {author} {\bibfnamefont {A.}~\bibnamefont {Udaltsov}}, \bibinfo {author} {\bibfnamefont {A.}~\bibnamefont {Duplinskiy}}, \bibinfo {author} {\bibfnamefont {V.}~\bibnamefont {Kurochkin}}, \ and\ \bibinfo {author} {\bibfnamefont {Y.}~\bibnamefont {Kurochkin}},\ }\bibfield  {title} {\enquote {\bibinfo {title} {Influence of chirp, jitter, and relaxation oscillations on probabilistic properties of laser pulse interference},}\ }\href@noop {} {\bibfield  {journal} {\bibinfo  {journal} {IEEE Journal of Quantum Electronics}\ }\textbf {\bibinfo {volume} {57}},\ \bibinfo {pages} {1--7} (\bibinfo {year} {2021})}\BibitemShut {NoStop}%
\end{thebibliography}
\providecommand{\noopsort}[1]{}\providecommand{\singleletter}[1]{#1}%

\end{document}